\documentclass[showpacs,aps,twocolumn,prl]{revtex4}

\usepackage{graphicx}
\usepackage{amssymb}

\def\fref#1{Fig.~\ref{#1}}
\renewcommand{\vec}{\mathbf}

\begin{document}

\title{Suppression of magnetic relaxation processes in hard superconductors
by a transverse AC magnetic field}

\author{L.~M. Fisher}
\author{A.~V. Kalinov}
\author{I.~F. Voloshin}
\affiliation{All-Russian Electrical Engineering Institute, 12 Krasnokazarmennaya Street, 111250
Moscow, Russia}
\author{V.~A. Yampol'skii}
\affiliation{Institute for Radiophysics and Electronics NASU, 12 Proskura Street, 61085 Kharkov,
Ukraine}

\begin{abstract}
The effect of the transverse AC magnetic field on relaxation process in YBa$_2$Cu$_3$O$_x$
melt-textured superconductor was studied. A factor of 50 suppression of the relaxation rate could
be achieved at the expense of some reduction in the maximum trapped field, with the
magnetic-induction gradient being unchanged. This phenomenon is interpreted as a result of an
increase of the pinning force after the action of the transverse AC magnetic field.
\end{abstract}

\date{\today}

\pacs{74.25.Nf, 74.25.Qt}

\maketitle

The electrodynamics of hard superconductors was under study for a long time. A wide variety of
interesting phenomena were discovered and interpreted on the basis of the developed concepts. One
of them is the suppression of static magnetization under the action of a transverse AC magnetic
field. First, this phenomenon was observed and interpreted by Yamafuji and coauthors in
Refs.~\onlinecite{yama1,yama2}. If one places a superconducting strip into a perpendicular static
(DC) magnetic field ${\vec H}$ and then apply an AC magnetic field ${\vec h}(t)$ along the strip
surface, the critical profile of the static magnetic flux changes noticeably. The short vortices
oriented in perpendicular to the strip are shown to bend and move in such a way that the static
magnetic flux tends to become homogeneous. This phenomenon and its consequences were studied
theoretically in detail in recent papers \cite{brand1, brand2}.

The other origin of the suppression of the static magnetization by the transverse AC magnetic field
(the collapse phenomenon) was considered in Refs. \onlinecite{col-new, cut, kin, cross, felpap,
aniscol}. If the vortex length $L$ exceeds significantly the penetration depth of the AC field, the
vortex bend does not play an essential role and the flux-line cutting mechanism (the phenomenon
predicted by Clem \cite{Clem}) is put in the forefront. As was shown in Refs.~\onlinecite{cut, kin,
felpap}, it is the flux-line cutting that provides the homogenization of the static magnetic flux
in the wide areas of the sample bulk where the AC field has penetrated (the creation of collapse
zone). If the AC amplitude $h$ exceeds the penetration field $H_p$ the distribution of the static
magnetic induction becomes homogeneous in the whole sample bulk and the magnetic moment disappears.
The collapse of the magnetization, closely related to the collapse of the transport current
\cite{collapse, buf}, was observed in Refs.~\onlinecite{park, col-new, felpap, aniscol}.

It is well-known that the inhomogeneous magnetic flux distribution is metastable. According to the
classical paper by Anderson \cite{Ander}, such a state of a superconductor relaxes to the
homogeneous one following the logarithmic law. Within the existing concept on the collapse, one
could expect a noticeable decrease of the relaxation rate of the magnetic moment after the action
of the AC field. Indeed, the surface homogeneous region of the sample, the collapse zone, should be
filled by the vortices and the critical gradient should be established before the vortices start to
leave the sample and the magnetic moment begins to decrease. In the present paper we have checked
this assumption and made sure that the relaxation rate of the DC magnetic moment is actually
decreased significantly by the action of the transverse AC magnetic field. Surprisingly, we have
observed a striking concomitant phenomenon which appears to be of general interest in vortex
matter. Not only the magnetic moment does not changes, but the spatial distribution of the static
magnetic flux holds the shape without significant relaxation for a long time after the action of AC
field.

A melt-textured YBCO plate-like sample of $9.3\times7.4\times1.5$~mm$^3$ in sizes was cut from a
homogeneous part of a bulk textured cylinder grown by the seeding technique. The homogeneity was
checked by a scanning Hall-probe. The \textbf{c}-axis was perpendicular to the largest face of the
sample. The characteristic value of the critical current density $J_c$ in the \textbf{ab} plane is
of the order of 13~kA/cm$^2$. The `static' magnetization $M$ was measured by a vibrating sample
magnetometer in the external magnetic field $\mathbf{H}\|\mathbf{c}$ created by an electromagnet.
The zero-field cooled sample was exposed to the magnetic field of 12~kOe which was further reduced
to 5~kOe; from this point, the relaxation measurement was started. The above-mentioned fields are
essentially higher than the penetration field of the sample (1.6~kOe). A commercial Hall-probe was
fixed on the sample to measure the evolution of the magnetic induction locally. The other
Hall-probe with the sensitive zone of $0.3\times0.3$~mm$^2$ was used to scan the magnetic induction
distribution on the sample surface. The distance from the sample to the sensitive zone of the probe
was about 0.2~mm. The Hall measurements were performed in the zero field after exposition of the
sample to $H=12$~kOe. The AC magnetic field $\mathbf{h}\|\mathbf{ab}$ was a computer-generated
triangle-wave with the frequency $F=140$~Hz. The AC field could be applied for a definite number of
periods and stopped exactly at the end of the period (at $h=0$). In our experiments we used 999
full cycles of the AC field. All the measurements were performed at the liquid nitrogen temperature
$T=77$~K.

\begin{figure}[tb]
\includegraphics[width=0.45\textwidth,height=0.4\textwidth]{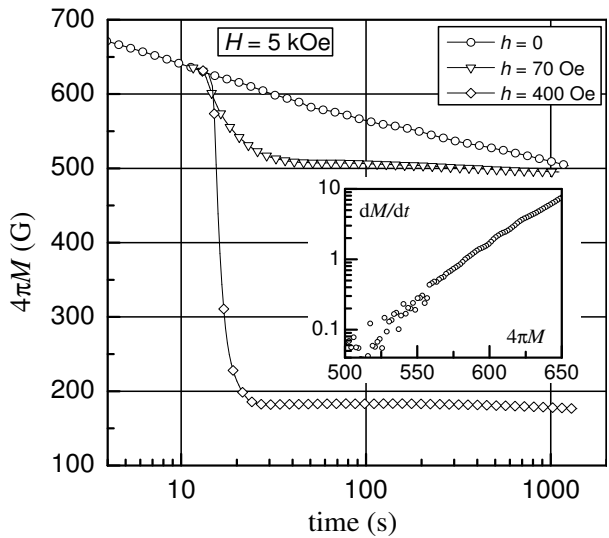}
\caption{\label{M-H} Relaxation of the magnetization without ({\Large $\circ$}) and with the action
of the AC field ($\triangledown$, $\diamond$). Inset shows the dependence of $\mathrm d M/\mathrm d
t \propto E$ on $M\propto J$.}
\end{figure}

Figure \ref{M-H} demonstrates the influence of the orthogonal AC magnetic field on the relaxation
of $M$. The conventional (without the AC field) relaxation is shown to follow the logarithmic law,
that implies an exponential current-voltage characteristic (CVC) $E \propto
\exp[(-U/kT)(1-J/J_{c0})]$ \cite{Ander} (see inset), where $J_{c0}$ is a depinning current at the
zero temperature and $U$ is a pinning-well depth. For the other run, at $t\approx 20$~s the AC
field of the amplitude $h$ was applied, followed by the sharp drop in the magnetization due to the
collapse-effect \cite{cross}. When the AC field action has finished, we can see almost no
relaxation for the first 100 seconds and essentially reduced logarithmic relaxation rate $S =
\mathrm d M/ \mathrm d \ln t$ for the rest of the observation period.

\begin{figure}[tb]
\includegraphics[width=0.45\textwidth,height=0.4\textwidth]{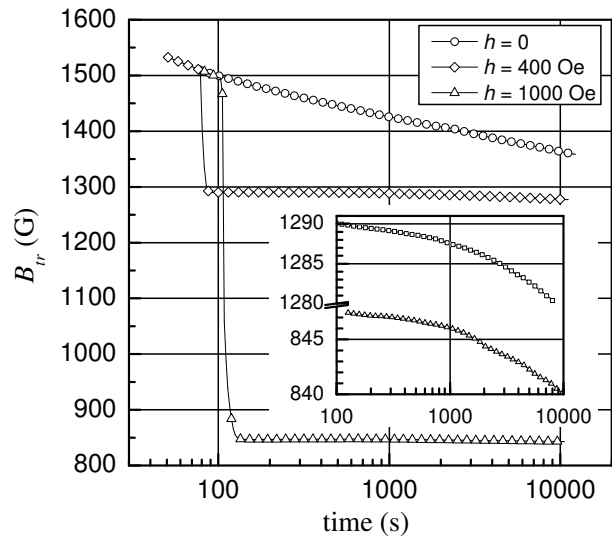}
\caption{\label{B-t} The influence of the AC field on the relaxation of the normal component of the
trapped magnetic induction $B_{tr}$ in the center of the largest face of the sample without
({\Large $\circ$}) and with the action of the AC field ($\diamond$, $\vartriangle$). Inset shows
the tails of suppressed relaxation in enlarged scale.}
\end{figure}

To reduce the minor effect of the magnetic-field relaxation in the electromagnet on the
magnetization we studied the temporal dependence of the trapped magnetic induction $B_{tr}$ in the
central part of the sample without any external magnetic field by the Hall-probe. Three curves in
\fref{B-t} correspond to the same magnetic prehistory of the sample, but differ from each other
because of the action of the AC field with different amplitudes. The upper curve is obtained for
the case when the AC field was not switched on during the measurement at all, $h=0$. Two lower
curves demonstrate the influence of the AC field on the relaxation of $B_{tr}$. All curves in
\fref{B-t} follow the same logarithmic-like law before switching on the AC field. Switching on the
AC field results in a giant suppression of the relaxation rate (5 -- 60 times). The inset to
\fref{B-t} shows that the decrease of the relaxation rate depends essentially on the amplitude of
the AC field and $B_{tr}$ does not follow the logarithmic law for the whole time-window.

The above mentioned results seem clear enough within the existing conception of the collapse.
Indeed, the abrupt decrease of the static magnetization under the action of the AC field (the jumps
shown in \fref{M-H} and \ref{B-t}) is connected to the suppression of the DC shielding currents in
all sample regions where the AC field penetrates. There exist two different mechanisms of such a
suppression which are actual in the system of relatively short vortices \cite{yama1, yama2} and
long ones \cite{col-new, cut, cross, felpap}. In our case, the both mechanisms play perhaps the
essential role. The suppression of the relaxation of $M$ and $B_{tr}$ could be easily interpreted
as a result of the collapse: vortices can not leave the sample before the sufficient gradient of
the magnetic induction would be restored in the surface regions of the sample. From the macroscopic
point of view, the other explanation looks plausible. As we reduce the magnetization (and current
density) significantly, we should get a huge decrease in $\mathrm d M/\mathrm d t\propto E$ in
accordance with the exponential CVC (see inset to \fref{M-H}). However, this explanation will work
only in the case of the uniform distribution of the shielding currents.

\begin{figure}[tb]
\includegraphics[width=0.45\textwidth,height=0.4\textwidth]{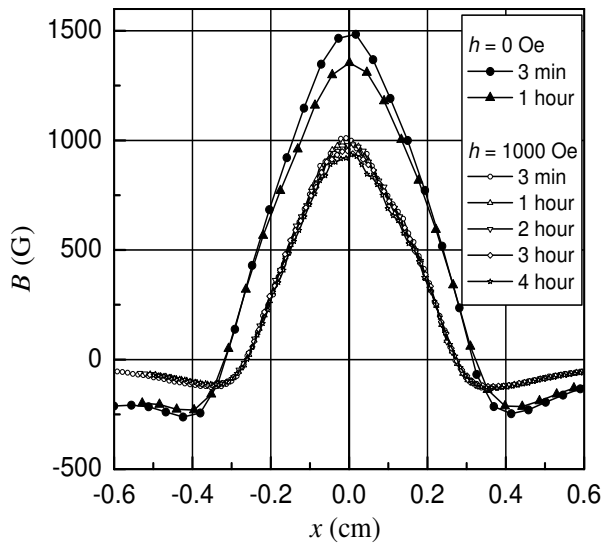}
\caption{\label{B-x} Relaxation of the spatial distribution of the magnetic induction on the sample
surface without (closed symbols) and after the action of the AC field (open symbols).}
\end{figure}

\begin{figure}[tb]
\includegraphics[width=0.45\textwidth,height=0.6\textwidth]{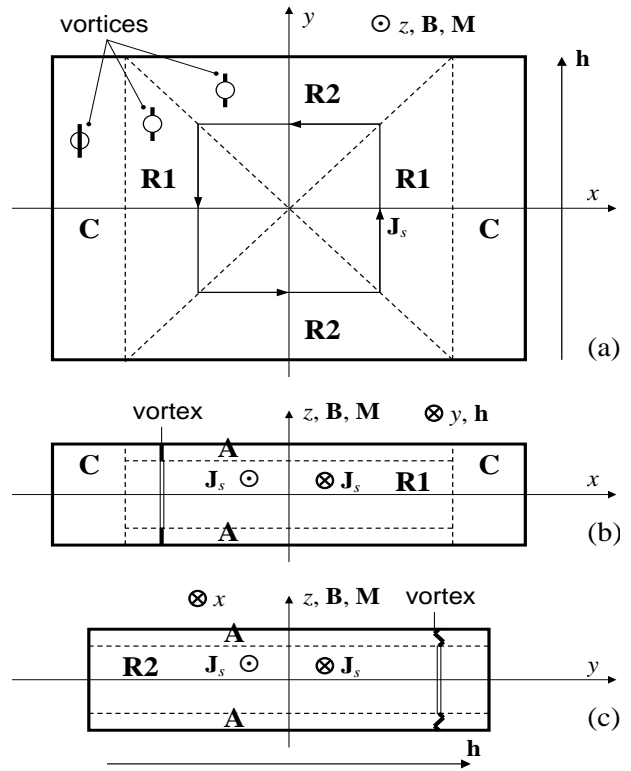}
\caption{\label{sketch} Schematic view of the cross-sections of the sample. Regions marked with `C'
are fully penetrated by the AC field, there is no shielding current ($J_s$) in the $xy$-plane here.
The vortices in regions `R1' and `R2' are partially disturbed (bent or muddled in the $zy$-plane).
This disturbance occurs in the `A'-regions corresponding to the AC-field penetration depth along
the $z$-axis.}
\end{figure}

To clarify the real influence of the AC field on the relaxation process we have investigated the
temporal evolution of the DC magnetic field distribution by the scanning Hall-probe. The spatial
distribution $B(x)$ (across the AC field direction) of the trapped DC magnetic field is shown in
\fref{B-x}. The unperturbed curves (closed symbols) demonstrate Bean-like profiles of the magnetic
flux distribution which relaxes obviously in one hour. After the action of the AC field (open
symbols) the relaxation is suppressed significantly in accordance with the data of \fref{B-t}. The
change in the profile width along $x$-axis is due to the collapse of the DC shielding currents in
the regions of the sample where the AC field has penetrated (the `collapsed' regions marked with
`C' in \fref{sketch}). Even at a first glance at the curves in \fref{B-x} one can reveal a very
important consequence of the action of the transverse AC field. The magnetic field distribution in
the central part of the sample before and after the action of the AC field differs by the vertical
displacement, with the gradient being the same. This means that the current densities remain their
initial values in all sample regions where the AC field has not penetrated along the $x$-direction.
So, after the action of the AC field, we get the nonuniform situation then the shielding-current
density remains changeless or becomes zero.

The most interesting feature of the relaxation after the action of the AC field is the following.
Not only the magnetization and the magnetic induction but the distribution of the DC-field
gradient, i.\ e.\ distribution of the DC shielding currents, stop the relaxation after the action
of the AC field. The vortices in the central part of the sample (`R1' and `R2' regions in
\fref{sketch}) could not move into the collapsed regions (`C'). This is probably due to
impossibility to overcome the cutting barrier \cite{Indenbom}, because the `C'-regions are (at
least partially) filled with the vortices aligned with the $y$-axis. However this is not enough to
stop the relaxation at all because there are no visible obstacles for vortices in the `R2'-region
to creep along the $y$-direction.

We suppose that the penetration of the AC field from the largest ($xy$) faces of the sample is of
great importance. The first possible reason is an increase in the vortex length accompanied by the
pinning energy gain, tending to the increase of the critical current density, because the vortices
turn out to be `anchored' by their tails (in the `A'-regions). The elongation of the vortices
occurs in such a way (along $y$-axis) that there is no additional Lorentz force directed outside
the sample (along $x$- and $y$-directions). Thus, the shielding currents become subcritical
resulting in the exponential decrease of the relaxation rate. However, to explain the significant
decrease of $S$ by the vortex-length increase along, the relative vortex elongation must be of the
order of the logarithmic relaxation-rate decrease, which is hardly possible. Moreover, in the
collective pinning approach \cite{Blatter} the vortex elongation gives only sublinear term in the
free energy. The other explanation could be based on the anisotropy of the pinning force in the
YBCO superconductors. After the action of the AC field, parts (segments) of the vortices in the
`A'-regions can be aligned and locked-in in the \textbf{ab}-plane leading to an increase of the
pinning force. The action of the AC field could also result in the bend or muddling of the vortices
in the `A'-regions at the scale of 100--1000 vortex-lattice periods. As a result, a significant
increase of the vortex-bundle size $R$ can take place. This, in turn, can lead to the increase of
the collective pinning potential \cite{Blatter} and to the observed suppression of the relaxation
rate.

Summarizing, we have observed the giant decrease of the relaxation rate after the action of the
transverse AC magnetic field. This effect exists in the relaxation of the `static' magnetization
and the local value of the trapped magnetic induction. Besides, after the action of the AC magnetic
field, the distribution of the magnetic induction on the sample surface turns out to be frozen with
the same critical current density. The observed effect can be a result of the increase of the
collective pinning potential due to the increase of the vortex-bundles dimension in the
\textbf{ab}-plane. The increase of the vortex length accompanied by partial locking of the vortex
segments in the \textbf{ab}-planes could also be a reason of this phenomenon.

We thank A. L. Rakhmanov for helpful discussions. This work is supported by INTAS (grant 01-2282),
RFBR (grant 03-02-17169), and Russian National Program on Superconductivity
(contract~40.012.1.1.11.46).

\end{document}